\documentclass[twocolumn,amsmath,amssymb,pre]{revtex4}
\usepackage{amsmath}
\usepackage{amsfonts}
\usepackage{amssymb}
\usepackage{graphicx}
\usepackage{epstopdf}
\usepackage{bbm}
\usepackage{amssymb}
\usepackage{esint}

\newcommand{\be}{\begin{equation}}
\newcommand{\ee}{\end{equation}}
\newcommand{\bey}{\begin{eqnarray}}
\newcommand{\eey}{\end{eqnarray}}
\newcommand{\bw}{\begin{widetext}}
\newcommand{\ew}{\end{widetext}}

\newcommand{\ra}{\rangle}
\newcommand{\la}{\langle}

\newcommand{\ba}{\begin{array}}
\newcommand{\ea}{\end{array}}
\newcommand{\bi}{\begin{itemize}}
\newcommand{\ei}{\end{itemize}}
\newcommand{\bem}{\begin{enumerate}}
\newcommand{\eem}{\end{enumerate}}

\makeatother

\begin{document}

\title{Characterization of random features of chaotic eigenfunctions in unperturbed basis}

 \author{Jiaozi Wang \footnote{ Email address: wangjz@ustc.edu.cn} and Wen-ge Wang
 \footnote{ Email address: wgwang@ustc.edu.cn}
  }

\affiliation{Department of Modern Physics, University of Science and Technology of China,
Hefei, 230026, China}

\date{\today}

\begin{abstract}
 In this paper, we study random features manifested in components of energy eigenfunctions
 of quantum chaotic systems, given in the basis of unperturbed, integrable systems.
 Based on semiclassical analysis, particularly on Berry's conjecture,
 it is shown that the components in classically allowed
 regions can be regarded as Gaussian random numbers in certain sense,
 when appropriately rescaled with respect to the average shape of the eigenfunctions.
 This suggests that, when a perturbed system changes from integrable to chaotic,
 deviation of the distribution of rescaled components in classically allowed regions
 from the Gaussian distribution may be employed as a measure for the ``distance'' to quantum chaos.
 Numerical simulations performed in the LMG model and the Dicke model show
 that this deviation coincides with the deviation of the nearest-level-spacing distribution
 from the prediction of random-matrix theory.
 Similar numerical results are also obtained in two models without classical counterpart.
\end{abstract}


\maketitle


\section{Introduction}

 A commonsense in the field of quantum chaos is that energy eigenfunctions (EFs) of chaotic systems
 should show certain random feature \cite{Haake,CC94book,Berry77,Berry91},
 though their Hamiltonian matrices are deterministic and some of them even may show a sparse structure.
 This property has vast applications in various fields
 \cite{Stockmann,Meredith98,
 Iz96,Prosen03,pre-98, Gnutzmann10,Kaplan09,pre02-LMG,qd2,opt1,nuc3}.
 In particular, it is of relevance to thermalization
 \cite{Izlev11,Haake12,Deutsch91,Sr94,Rigol08,Rigol12,Gemmer15,ETH-review,ETH-Chaos},
 a topic which has attracted renewed interest in recent years.

 According to Berry's conjecture, for EFs of chaotic systems expressed in the configuration space,
 their components in classically allowed regions
 can be regarded as being given from certain Gaussian random numbers \cite{Berry77}.
 Based on this conjecture, it is natural to expect
 that, when expanded in the bases of unperturbed integrable systems, the EFs
 should show certain random feature as well within appropriate regions.
 Indeed, numerical simulations revealed such a feature for main bodies of EFs (see, e.g., Ref.\cite{Buch82}).
 However, more detailed study showed that the distribution of components of EFs
 usually exhibit notable deviation from the Gaussian distribution, which is
 predicted by the random-matrix theory (RMT) (see, e.g., Ref.\cite{Meredith98}).

 More recently, numerical simulations show that, if EFs are rescaled with respect to
 their average shape, the above-discussed deviation can be considerably reduced \cite{EFchaos-Benet03,EFchaos-WW}.
 This gives a clue to a solution to a long-standing problem in the field of quantum chaos,
 that is, in which way statistical properties of EFs may be employed
 to give a quantitative measure for the ``distance'' to chaos.
 The above-mentioned numerical simulations suggest that deviation of the distribution
 of rescaled components of EFs from the Gaussian distribution should be a candidate for such a measure
 of ``distance''.
 However, presently, the situation is not completely clear,
 because in some cases this measure shows notable deviations from results
 obtained from statistical properties of spectra, e.g., from deviation of the
 nearest-level-spacing distribution from the prediction of the RMT \cite{EFchaos-Benet03,EFchaos-WW}.

 In this paper, based on semiclassical analysis, particularly on the Berry's conjecture,
 we study random features manifested in components of EFs of chaotic
 systems in integrable bases.
 Our analysis shows that the distribution of the components in classically-allowed regions
 indeed should have a Gaussian form, under a rescaling procedure which is more appropriate than
 that adopted in Refs.\cite{EFchaos-Benet03,EFchaos-WW}.
 Our numerical simulations performed in the Lipkin-Meshkov-Glick (LMG) model and
 the  Dicke model show that, adopting this new rescaling procedure,
 deviation of the distribution of components from the Gaussian distribution
 coincides quite well with that obtained from the statistics of spectra.
 We also study some models without any classical counterpart and find similar results.

 The paper is organised as follows.
 In Sec.\ref{sect-chaosEF}, a detailed semiclassical analysis is carried out for
 random features manifested in components
 of EFs of chaotic systems in integrable bases.
 Numerical simulations in two models with classical counterparts
 are discussed in Sec.\ref{sect-num-Lip}.
 Then, in Sec.\ref{sect-num-Ising}, we discuss numerical simulations
 performed in two models without classical counterpart.
 Finally, conclusions and discussions are given in Sec.\ref{sect-Conclusion}.

\section{Random features of chaotic EFs}\label{sect-chaosEF}

 In Sec.\ref{sect-chaosEF-sa},  based on semiclassical
 analysis we discuss random features of chaotic EFs.
 Then, making use of results obtained, we discuss a quantitative characterization of
 the random feature in Sec.\ref{sect-measure}.

\subsection{Semiclassical analysis of chaotic EFs}\label{sect-chaosEF-sa}

 Consider a quantum system, which has an $f$-dimensional classical counterpart, with a Hamiltonian
\begin{equation}\label{H}
H=H_0+\lambda V,
\end{equation}
 where $H_0$ indicates the Hamiltonian of an integrable system and $V$ is a perturbation.
 Within certain regime of the parameter $\lambda$,
 the classical counterpart of the system $H$ undergoes a chaotic motion.
 In this section, we consider a chaotic system $H$.
 In terms of action-angle variables, $H_0$ is written as
\be
H_0=\boldsymbol{d}\cdot \boldsymbol{I} +c_0,
\ee
 where $\boldsymbol{I}=(I_1,I_2,\cdots,I_f)$ is the action variable, $\boldsymbol{d }$
 is a parameter vector, $\boldsymbol{d }=(d _{1},d _{2},\cdots,d _{f})$, and $c_0$ is a constant parameter.

 In the quantum case, we use $|\boldsymbol n\ra$ to denote the eigenbasis of $\boldsymbol{I}$,
 with $\boldsymbol I |\boldsymbol n\ra = \boldsymbol {I_n}|\boldsymbol n\ra$,
 where $\boldsymbol{n}=(n_1,n_2,\cdots,n_f)$ is an integer vector and $\boldsymbol{I_n}=\boldsymbol n \hbar$.
 The Hamiltonian $H_0$ has a diagonal form in this basis with eigenvalues denoted by $E_{\boldsymbol{n}}^{0}$,
\begin{equation}
H_{0}|\boldsymbol{n}\rangle=E_{\boldsymbol{n}}^{0}|\boldsymbol{n}\rangle.
\end{equation}
We use $|E_\alpha\rangle$ to denote eigenstates of $H$ with eigenvalues $E_\alpha$ in energy order,
\be
{H}|E_\alpha \rangle =E_\alpha |E_\alpha\rangle.
\ee
 The expansion of $|E_\alpha\ra$ in the basis $|\boldsymbol{n}\rangle$ is written as
\be
|E_{\alpha}\rangle=\sum_{\boldsymbol{n}}C_{\alpha\boldsymbol{n}}|\boldsymbol{n}\rangle,
\ee
 with $C_{\alpha\boldsymbol{n}} = \la \boldsymbol{n}|E_\alpha\ra$.
 Below in this section, we discuss {random features} manifested in
 the components $C_{\alpha\boldsymbol{n}}$ and their statistical
 properties in chaotic systems.

 In terms of the wave functions of $|E_\alpha\rangle$ and $|\boldsymbol{n}\rangle$ in the momentum space,
 denoted by $\psi_{\alpha}(\boldsymbol{p})$ and  $\psi_{\boldsymbol{n}}^{0}(\boldsymbol{p})$, respectively,
 the components $C_{\alpha \boldsymbol{n}}$ are written as
\be\label{Calpha-int}
C_{\alpha\boldsymbol{n}}=\int(\psi_{\boldsymbol{n}}^{0}(\boldsymbol{p}))^{*}\psi_{\alpha}(\boldsymbol{p})d\boldsymbol{p}.
\ee
 Generically, a wave function $\psi_{\alpha}(\boldsymbol{p})$ can be written in the following form,
\begin{equation}
\psi_{\alpha}(\boldsymbol{p})=A_{\alpha}(\boldsymbol{p})\sqrt{\Pi_{\alpha}(\boldsymbol{p})},\label{eq-psiP-chaos}
\end{equation}
 where $\Pi_{\alpha}(\boldsymbol{p})$ indicates local average of
 $|\psi_\alpha (\boldsymbol{p})|^2$.
 Then, $C_{\alpha\boldsymbol{n}}$ is written as
\begin{gather}
 C_{\alpha\boldsymbol{n}}
 =\int A_{\alpha}(\boldsymbol{p})(\psi_{\boldsymbol{n}}^{0}(\boldsymbol{p}))^*
\sqrt{\Pi_{\alpha}(\boldsymbol{p})}d\boldsymbol{p}. \label{C-alpha-n}
\end{gather}
 According to the Berry's conjecture \cite{Berry77}, in a chaotic system the quantity $A_\alpha (\boldsymbol{p})$
 should have random phases.
 This implies that the components $C_{\alpha\boldsymbol{n}}$ can be effectively regarded as some random numbers.

 In realistic physical models,
 the average shape of $|C_{\alpha\boldsymbol{n}}|^2$
 is usually not uniform with respect to the perturbed and unperturbed energies.
 Due to this nonuniformity, the statistical distribution of the components $C_{\alpha\boldsymbol{n}}$
 can not have a Gaussian shape \cite{Meredith98}.
 But, if they are rescaled such that the effect of average shape of EFs is appropriately taken into account,
 it should be possible for their distribution to have a Gaussian form.
 Below, we derive a semiclassical expression for the
 average shape of $|C_{\alpha\boldsymbol{n}}|^2$ that is suitable for this purpose.

 The Wigner function supplies a useful tool in semiclassical analysis of eigenstates.
 We use $\psi_{\alpha}(\boldsymbol{r})$ and  $\psi_{\boldsymbol{n}}^{0}(\boldsymbol{r})$ to indicate
 the wave functions of $|E_\alpha\rangle$ and $|\boldsymbol{n}\rangle$ in the coordinate space, respectively.
 The Wigner function corresponding to $\psi_{\alpha}(\boldsymbol{r})$,
 denoted by $W_{\alpha}(\boldsymbol{p},\boldsymbol{q})$, is written as
\be
W_{\alpha}(\boldsymbol{p,q})=\frac{1}{(2\pi\hbar)^{f}}\int_{-\infty}^{\infty}\psi_{\alpha}^{*}
(\boldsymbol{q}+\frac{\boldsymbol{r}}{2})\psi_{\alpha}(\boldsymbol{q}-\frac{\boldsymbol{r}}{2})
e^{i\boldsymbol{p\cdot r}/\hbar}d\boldsymbol{r},
\ee
 and similar for the Wigner function corresponding to $\psi_{\boldsymbol{n}}^{0}(\boldsymbol{r})$,
 denoted by $W_{\boldsymbol{n}}^{0}(\boldsymbol{p},\boldsymbol{q})$.
 As is known, in a chaotic system, the averaged Wigner function, with average taken within
 certain small regions of the phase space, has the following expression,
 \cite{Berry77,Berry91,Voros76,Voros77}
\be\label{eq-WFchaos}
\overline{W}_{\alpha}(\boldsymbol{p},\boldsymbol{q})
=\frac{\delta(H(\boldsymbol{p},\boldsymbol{q})-E_{\alpha})}{S({E_{\alpha}})},
\ee
where $S(E)$ represents the area of an energy surface with $H(\boldsymbol{p},\boldsymbol{q})=E$,
\be
S({E})=\int d\boldsymbol{p}d\boldsymbol{q}\delta(E-H(\boldsymbol{p},\boldsymbol{q})).
\ee
 Equation (\ref{eq-WFchaos}) gives that
\be
\Pi_{\alpha}(\boldsymbol{p})=\frac{1}{S({E_{\alpha}})}\int\delta(E_{\alpha}-H(\boldsymbol{p},
\boldsymbol{q}))d\boldsymbol{q}.
\ee

 Equation (\ref{eq-WFchaos}) implies that most eigenstates within a narrow
 energy window in a chaotic system should have close shapes.
 Therefore, when computing the average shape of $|C_{\alpha\boldsymbol{n}}|^{2}$
 for the purpose discussed above,
 one may perform an average within such a narrow energy window.
 For the convenience in discussion, we write a coarse-grained $\delta$-function as $\delta_\epsilon (E)$,
\be
\delta_{\epsilon}(E)=\begin{cases}
\frac{1}{\epsilon} & E\in[-\frac{\epsilon}{2},\frac{\epsilon}{2}],\\
0 & {\rm otherwise},
\end{cases}
\ee
 where $\epsilon$ is a small parameter.
 The choice of energy window $\epsilon$ should satisfy the following requirements:
 It is small in the classical case such that the energy surface almost does not change
 within the window,
 while, it is sufficiently large in the quantum case such that many energy levels
 are included within the window.

 Then, the average shape of EFs, denoted by $\langle|C_{\alpha\boldsymbol{n}}|^{2}\rangle$, is computed by
\be\label{eq-C2AV}
\langle|C_{\alpha\boldsymbol{n}}|^{2}\rangle=\frac{1}{N_{E_{\alpha}}}
\sum_{\alpha'}|C_{\alpha'\boldsymbol{n}}|^{2}
\delta_{\epsilon}(E_{\alpha'}-E_{\alpha}),
\ee
where
\be\label{eq-NEA}
N_{E_{\alpha}}=\sum_{\alpha{'}}\delta_{\epsilon}(E_{\alpha{'}}-E_{\alpha}).
\ee

 In order to derive an explicit expression for $\langle|C_{\alpha\boldsymbol{n}}|^{2}\rangle$,
 we make use of the following well-known expression of $|C_{\alpha\boldsymbol{n}}|^{2}$,
\begin{equation}
|C_{\alpha\boldsymbol{n}}|^{2}=(2\pi\hbar)^{f}\int d\boldsymbol{p}d\boldsymbol{q}W_{\alpha}
(\boldsymbol{p},\boldsymbol{q})W_{\boldsymbol{n}}^{0}(\boldsymbol{p},\boldsymbol{q}).\label{eq-wigner-wave}
\end{equation}
 Let us divide the phase space into small cells, denoted by $c_\sigma$ with a label $\sigma$,
 each having a volume $\delta \Omega$, meanwhile, keep
 the ratio $\delta \Omega /\hbar^f$ large such that there are many quantum
 states ``lying'' within each small cell.
 Then, $|C_{\alpha \boldsymbol{n}}|^2$ is written as
\be\label{eq-wwc}
|C_{\alpha\boldsymbol{n}}|^{2}=(2\pi\hbar)^{f}\sum_{\sigma}\int_{c_{\sigma}} d\boldsymbol{p}d\boldsymbol{q}
W_{\alpha}(\boldsymbol{p},\boldsymbol{q})W_{\boldsymbol{n}}^{0}(\boldsymbol{p},\boldsymbol{q}).
\ee
 Substituting Eq.(\ref{eq-wwc}) into Eq.(\ref{eq-C2AV})
 and performing the summation over the perturbed states $|E_{\alpha'}\ra$, one gets that
\be\label{eq-CAV}
 \la |C_{\alpha\boldsymbol{n}}|^{2} \ra =(2\pi\hbar)^{f}\sum_{\sigma}\int_{c_{\sigma}} d\boldsymbol{p}d\boldsymbol{q}
\langle{W}_{\alpha}(\boldsymbol{p},\boldsymbol{q})\rangle W_{\boldsymbol{n}}^{0}
(\boldsymbol{p},\boldsymbol{q}),
\ee
 where $\langle{W}_{\alpha}(\boldsymbol{p},\boldsymbol{q})\rangle$ indicates
 the average of ${W}_{\alpha}(\boldsymbol{p},\boldsymbol{q})$
 over perturbed states within a small energy window $\epsilon$.
 As  many energy levels are included within the window $\epsilon$,
 $\langle{W}_{\alpha}(\boldsymbol{p},\boldsymbol{q})\rangle$
 should vary slowly  within each small cell $c_\sigma$, such that
\be
\langle{W}_{\alpha}(\boldsymbol{p},\boldsymbol{q})\rangle \simeq \langle{W}_{\alpha}
(\boldsymbol{p}_{\sigma},\boldsymbol{q}_{\sigma})\rangle
\quad \text{ for }(\boldsymbol{p},\boldsymbol{q})\in c_{\sigma},
\ee
 where $(\boldsymbol{q}_\sigma,\boldsymbol{p}_\sigma)$ indicates the center of $c_\sigma$.
 Then, Eq.(\ref{eq-CAV}) gives that
\be\label{avC-WW}
 \la |C_{\alpha\boldsymbol{n}}|^{2}\ra \simeq (2\pi\hbar)^{f}\sum_{\sigma}\langle{W}_{\alpha}
(\boldsymbol{p}_{\sigma},\boldsymbol{q}_{\sigma})\rangle \overline{W}^0_{\boldsymbol{n}}
(\boldsymbol{p}_{\sigma},\boldsymbol{q}_{\sigma})\delta\Omega,
\ee
 where $\overline{W}^0_{\boldsymbol{n}}(\boldsymbol{p}_{\sigma},\boldsymbol{q}_{\sigma})$
 represents the average of the Wigner function of the integrable system
 within the cell $c_\sigma$,
\be
\overline{W}^0_{\boldsymbol{n}}(\boldsymbol{p}_{\sigma},\boldsymbol{q}_{\sigma})
=\frac{1}{\delta\Omega}\int_{c_{\sigma}}W_{\boldsymbol{n}}^{0}(\boldsymbol{p},\boldsymbol{q})
d\boldsymbol{p}d\boldsymbol{q}.
\ee

 Due to the classical smallness and quantum mechanical
 largeness of the energy windows $\epsilon$ discussed previously,
 $\langle{W}_{\alpha}(\boldsymbol{p},\boldsymbol{q})\rangle$ in Eq.(\ref{eq-CAV})
 obeys Eq.(\ref{eq-WFchaos}) in an approximate way, with
\be\label{eq-WAE}
\langle W_{\alpha}(\boldsymbol{p},\boldsymbol{q})\rangle\simeq\overline{W}_{\alpha}(\boldsymbol{p},\boldsymbol{q}),
\ee
 and its dependence on $\epsilon$ can be neglected.
 It is known that \cite{Berry77}
\be\label{W0-int}
\overline{W}_{\boldsymbol{n}}^{0}(\boldsymbol{p},\boldsymbol{q})=\frac{\delta(\boldsymbol{I}
(\boldsymbol{p},\boldsymbol{q})-\boldsymbol{I_{n}})}{(2\pi)^{f}}.
\ee
 Substituting Eqs.(\ref{eq-WAE}) and (\ref{W0-int}) into Eq.(\ref{avC-WW})
 and noting that the smallness of the cells $c_\sigma$ enables one to change
 the summation over $\sigma$ back to the integration over phase space,
 one gets the following semiclassical expression,
\be\label{eq-EFshapecl}
\langle|C_{\alpha\boldsymbol{n}}|^{2}\rangle \simeq \hbar^{f}\ \Pi(E_{\alpha},\boldsymbol{I_{n}}),
\ee
 where
\begin{gather}\label{Pi-E-I}
 \Pi(E,\boldsymbol{I}) = \frac{S(E,\boldsymbol{I})}{S({E})},
 \\ \label{eq-ov}
S(E,\boldsymbol{I})=\int d\boldsymbol{p}d\boldsymbol{q}\delta(E-H(\boldsymbol{p},
\boldsymbol{q}))\delta(\boldsymbol{I}-\boldsymbol{I}(\boldsymbol{p},\boldsymbol{q})).
\end{gather}
 Here, $S(E,\boldsymbol{I})$ indicates the overlap of the energy
 surface of $H(\boldsymbol{p},
\boldsymbol{q}) =E$ and the torus of $\boldsymbol{I}(\boldsymbol{p},\boldsymbol{q}) =\boldsymbol{I}$.
 Since Eq.(\ref{eq-WFchaos}) works for classically allowed regions only,
 so does Eq.(\ref{eq-EFshapecl}).
 Sometimes, quantities like $\Pi(E,\boldsymbol{I})$ are called
 {classical analog} of averaged EFs \cite{EFchaos-Benet03,EFchaos-Benet00,EFchaos-Borgonovi98}.

 Finally, we consider rescaled components denoted by
 $R_{\alpha\boldsymbol{n}}$, defined by
\be\label{eq-C9}
 R_{\alpha\boldsymbol{n}}=\frac{C_{\alpha\boldsymbol{n}}}{\sqrt{\langle|C_{\alpha\boldsymbol{n}}|^{2}\rangle}}.
\ee
 Discussions given above show that this quantity $R_{\alpha\boldsymbol{n}}$
 can be regarded as a Gaussian random number with mean zero.
 Note that $\langle |R_{\alpha \boldsymbol{n}}|^2\rangle = 1$
 according to Eq.(\ref{eq-C9}).

\subsection{A measure for ``distance" to quantum chaos}\label{sect-measure}

 Let us use $f(R)$ to denote the distribution of $R_{\alpha\boldsymbol{n}}$.
 According to results given in the above section, for a chaotic system,
 $f(R)$ should have a Gaussian form, i.e.,
\be
f(R)=f_G (R),
\ee
where $f_G (R)$ is the Gaussian distribution,
\be
f_{G}(R)=\frac{1}{\sqrt{2\pi}}\exp(-R^{2}/2),
\ee
 In the RMT, the Gaussian distribution is predicted directly for components of EFs \cite{Haake}.
 But, for Hamiltonians in realistic models with chaotic classical
 counterparts, as discussed above,
 it is the distribution of the rescaled components $R_{\alpha\boldsymbol{n}}$
 that should have a Gaussian form.
 On the other hand, in a nearly integrable system, the quantity $A_\alpha (\boldsymbol{p})$
 on the rhs of Eq.(\ref{C-alpha-n}) does not have random phases and
 EFs with close energies may have quite different shapes.
 As a result, the distribution $f(R)$ in nearly integrable systems
 should usually show notable deviation from $f_G (R)$.

 The above discussions suggest that deviation of $f(R)$ from $f_G (R)$ may be employed as an
 {measure} for the ``distance'' to quantum chaos.
 In order to quantitatively characterize the deviation, one may consider a quantity $\Delta_{EF}$
 defined by
\be\label{eq-DEF}
\Delta_{EF}=\int|I_{f}(R)-I_{f_G}(R)|dR,
\ee
where $I_f(R)$ and $I_{f_G}(R)$ indicate the cumulative distributions of $f(R)$ and $f_G(R)$, respectively,
 e.g., $I_f(R) = \int_{-\infty}^R dr f(r)$.
 As is well known, cumulative distributions usually exhibit less fluctuations compared with the origin distributions.

\begin{figure}
\includegraphics[width=1.1\linewidth]{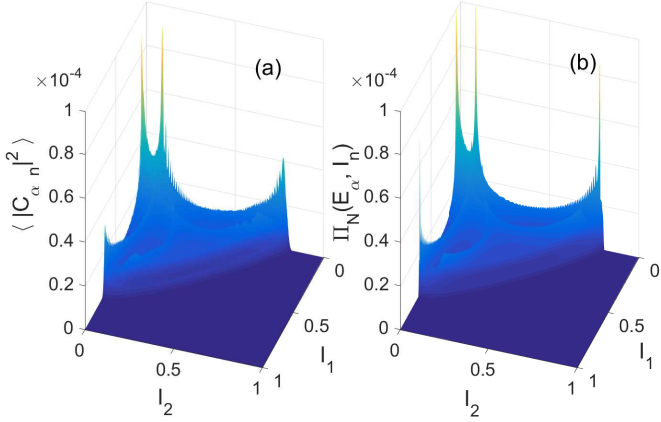}\caption{
(a): The average shape of EFs,
$\langle|C_{\alpha\boldsymbol{n}}|^{2}\rangle$,
 in the chaotic regime of the LMG model  with $\Omega=500$ and $\lambda=1$.
The average is taken over 500 EFs of $|E_\alpha\ra$ with $\epsilon=0.4175$.
(b): $\Pi_N({E_\alpha, \boldsymbol{I_n}})$, which is the normalized $\Pi(E_\alpha,\boldsymbol{I_n})$, as a
 classical analog of $\langle|C_{\alpha\boldsymbol{n}}|^{2}\rangle$ [see Eq.(\ref{eq-EFshapecl})].
}\label{fig-EF2D-LMG}
\end{figure}

 In the field of quantum chaos, the most-often used criterion for quantum chaos is
 given by statistical properties of spectra, e.g., by closeness of the nearest-level-spacing distribution $P(s)$
 to the prediction of RMT \cite{Haake}.
 It is known that the following distribution $P_W (s)$, which is obtained from Wigner's surmise,
\be
P_W (s) = \frac{\pi}{2}s\exp(-\frac{\pi}{4}s^{2}),
\ee
 gives a good approximation to the nearest-level-spacing distribution of the Gaussian
 orthogonal ensemble (GOE) in the large size limit.
 Quantitatively, the above-discussed closeness can be characterized by the following quantity $\Delta_E$,
\be\label{eq-DE}
\Delta_E =\int |I_P(s)-I_{P_W} (s)| ds,
\ee
 where $I_P(s)$ and $I_{P_W}(s)$ are cumulative distributions of $P(s)$ and $P_W(s)$, respectively.

\begin{figure}
\includegraphics[width=1.1\linewidth]{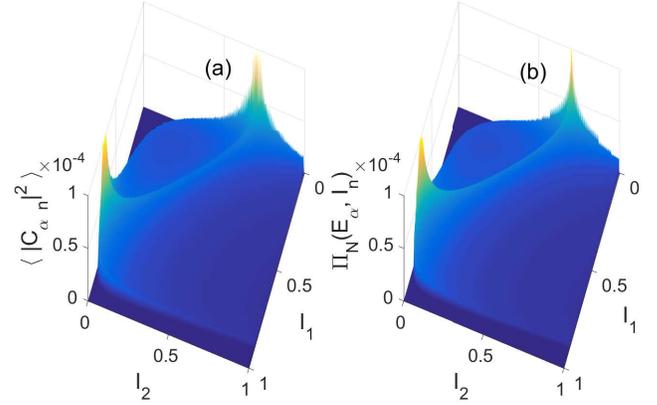}
\caption{Similar to Fig.\ref{fig-EF2D-LMG}, but for the Dicke model with $N=500$, $\lambda=1$,
 and $\epsilon=0.007$.}\label{fig-EF2D-Dicke}
\end{figure}

 In previous numerical simulations, deviation of the distribution of another rescaled components,
 denoted by $R_{\alpha \boldsymbol{n}}'$, from the Gaussian distribution
 was studied as a measure for the ``distance'' to chaos,
 where $R_{\alpha \boldsymbol{n}}'
 ={C_{\alpha \boldsymbol{n}}}/{\sqrt{\langle|C_{\alpha \boldsymbol{n}}|^{2}\rangle'}}$
 \cite{EFchaos-Benet03,EFchaos-WW,foot-region}.
 Here, in the computation of $\langle|C_{\alpha \boldsymbol{n}}|^{2}\rangle'$,
 in addition to an average over perturbed states with energies close to $E_\alpha$,
 a further average is taken over unperturbed states $|\boldsymbol{n}'\ra$ with
 unperturbed energies close to $E^0_{\boldsymbol{n}}$ by a small quantity $\epsilon_0$.
 Specifically,
\be
\langle |C_{\alpha\boldsymbol{n}}|^2\rangle'
=\sum_{\alpha',\boldsymbol{n}'} \frac{|C_{\alpha'\boldsymbol{n}'}|^{2}}{N_{E_{\alpha}}N_{E_{\boldsymbol{n}}}}
\delta_{\epsilon}(E_{\alpha'}-E_{\alpha})\delta_{\epsilon_{0}}(E_{\boldsymbol{n}'}-E_{\boldsymbol{n}}),
\ee
where $N_{E_\alpha}$ is defined in Eq.(\ref{eq-NEA}) and
\be
N_{E_{\boldsymbol{n}}}=\sum_{\boldsymbol{n}'}\delta_{\epsilon_{0}}(E_{\boldsymbol{n}'}-E_{\boldsymbol{n}}).
\ee
It was found that, in some cases  (not rare) in which the classical counterparts undergo chaotic motion
 and the distributions $P(s)$ are quite close to $P_W (s)$,
 the distributions of $R_{\alpha \boldsymbol{n}}'$, denoted by $g(R')$, deviate notably
 from the Gaussian distribution.

 In view of the semiclassical analysis given in the previous section,
 it is understandable that deviation of $g(R')$ from $f_G(R')$
 may be larger than that of $f(R)$ from $f_G(R)$.
 In fact, unperturbed basis states $|\boldsymbol{n}\rangle$ with close energies
 $E^0_{\boldsymbol{n}}$ may correspond to quite different values of $\boldsymbol{I}_{\boldsymbol{n}}$,
 meanwhile, according to Eq.(\ref{eq-EFshapecl}), the values of $\la |C_{\alpha \boldsymbol{n}}|^2\ra $
 of those $\boldsymbol{n}$, for which
 $\boldsymbol{I}_{\boldsymbol{n}}$ are far from each other, are usually quite different.
 As a result, taking average over unperturbed basis states with close $E^0_{\boldsymbol{n}}$
 may drive the distribution
 of rescaled components away from the Gaussian distribution.

 Therefore, in order to obtain rescaled components that have a Gaussian distribution,
 no average should be taken over the unperturbed energies  $E^0_{\boldsymbol{n}}$.
 We would remark that, when the torus of $\boldsymbol{I} = \boldsymbol{I}_{\boldsymbol{n}}$ does not change rapidly
 with $\boldsymbol{n}$, an average over a neighborhood of $\boldsymbol{n}$ is allowed.
 We did not mention this averaging procedure in the above discussions, because
 it is unnecessary in the derivation of Eq.(\ref{eq-EFshapecl}).

\section{Numerical simulations in two models with classical counterparts}\label{sect-num-Lip}

 In order to test the above analytical results, numerical simulations have been performed
 in two models possessing classical counterparts,  the LMG model and the Dicke model.
 In this section,  we first briefly discuss the two models, then, present numerical results
 about the distribution $f(R)$ and about the suggested ``distance'' to chaos, namely, $\Delta_{EF}$,
 in comparison with other ``distances''.

 \begin{figure}
\includegraphics[width=0.96\linewidth]{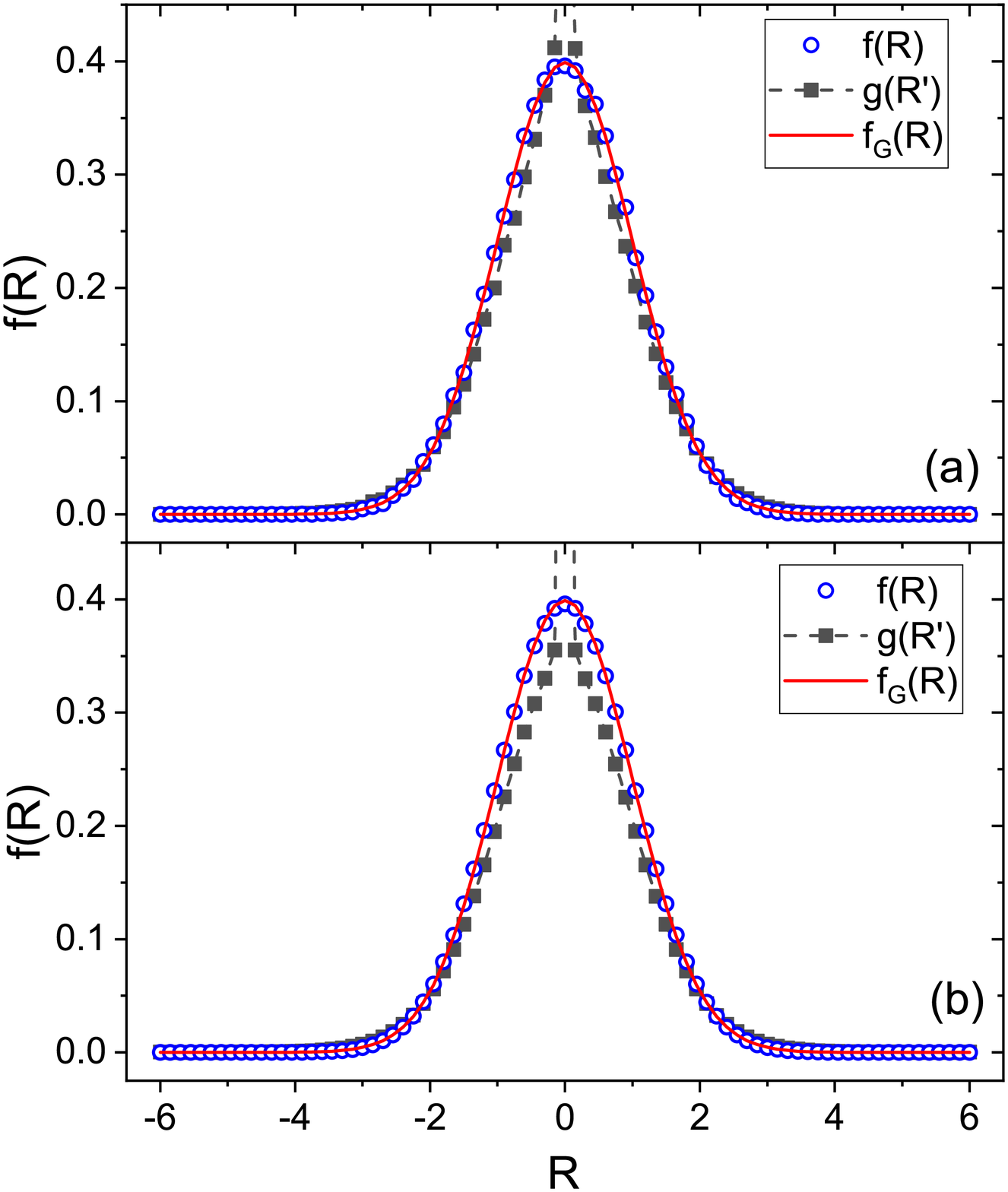}
\caption{The distribution $f(R)$ (open circles) and $g(R')$ (solid blocks with dashed lines)
for $\lambda=1$ in the LMG model with (a) $\Omega=80$ and (b) $\Omega=1000$.
 The solid curves indicate the Gaussian distribution $f_{G}(R)$.
}\label{DisLMG}
\end{figure}

\subsection{Models}

 \begin{figure}
\includegraphics[width=0.96\linewidth]{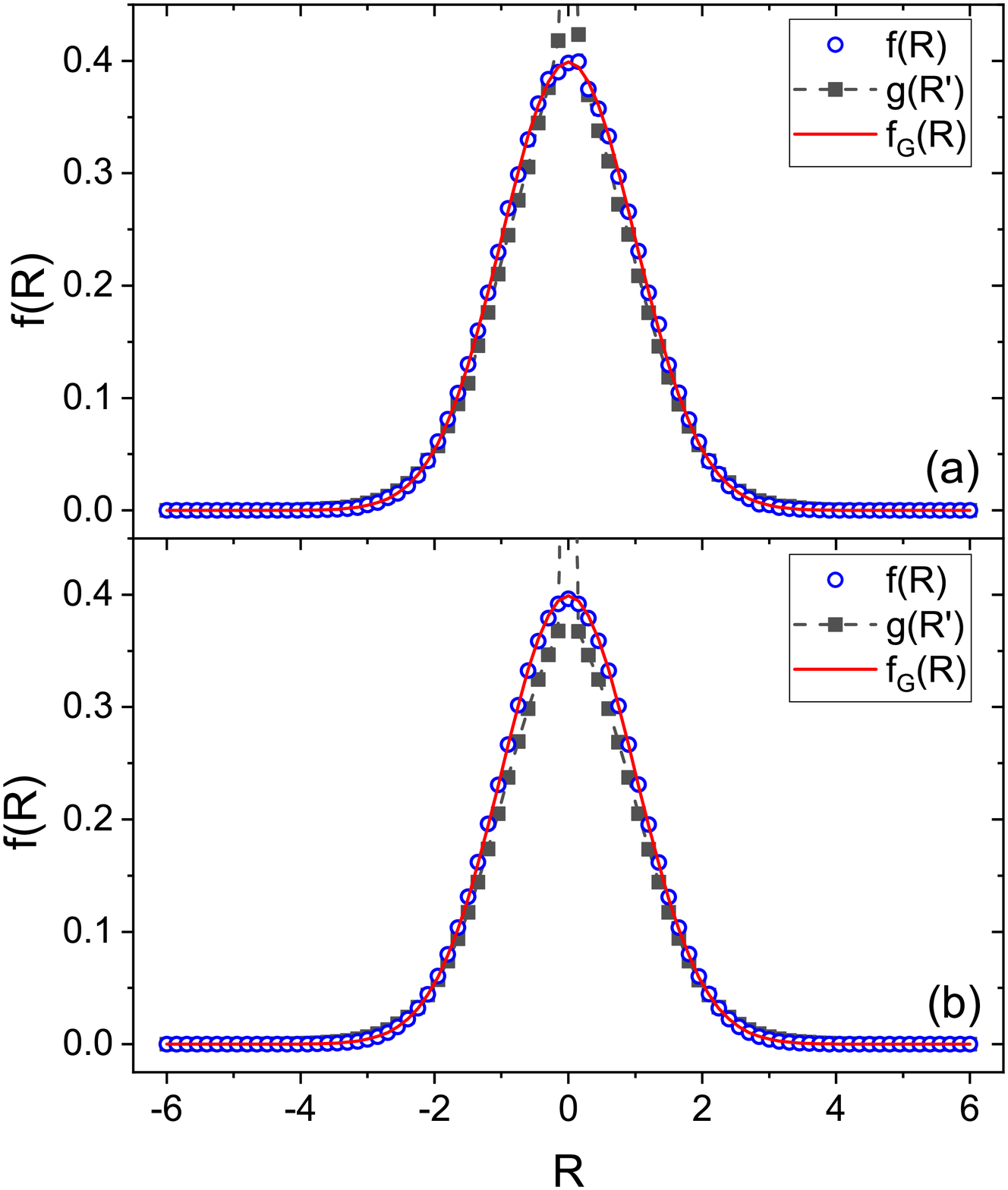}
\caption{Similar to Fig.\ref{DisLMG}, but for the Dicke model with $\lambda=1$, (a)
 $N=80$,  (b)$N=1000$.
}\label{DisDicke}
\end{figure}

The first model we employ is a three-orbital LMG model \cite{LMG}.
 This model is composed of $\Omega$ particles, occupying three energy levels labeled by $r=0,1,2$, each with
 $\Omega$-degeneracy.
 Here, we are interested in the collective motion of this model, for which the dimension of the
 Hilbert space is $\frac 12 (\Omega+1)(\Omega +2)$.
 We use $\epsilon_{r}$ to denote the energy of the $r$-th level
 and, for brevity, we set $\epsilon_{0}=0$.
 The Hamiltonian of the model,  in the form in Eq.(\ref{H}), has \cite{pre-98}
\begin{gather}
 H_{0}=\epsilon_{1}K_{11}+\epsilon_{2}K_{22},  \\
V=\sum_{t=1}^{4}\mu_{t}V^{(t)},
\end{gather}
where
\begin{eqnarray}
V^{(1)}=K_{10}K_{10}+K_{01}K_{01},\ V^{(2)}=K_{20}K_{20}+K_{02}K_{02},\nonumber \\
 V^{(3)}=K_{21}K_{20}+K_{02}K_{12},\ V^{(4)}=K_{12}K_{10}+K_{01}K_{21}. \ \
\end{eqnarray}
Here, the operators $K_{rs}$ are defined by
\be
K_{rs}=\sum_{\gamma=1}^{\Omega}a_{r\gamma}^{\dagger}a_{s\gamma},\quad r,s=0,1,2,
\ee
where $a^{\dagger}_{r\gamma}$ and $a_{r\gamma}$ are fermionic creation and annihilation
operators obeying the usual anti-commutation relations.

\begin{figure}
\includegraphics[width=1\linewidth]{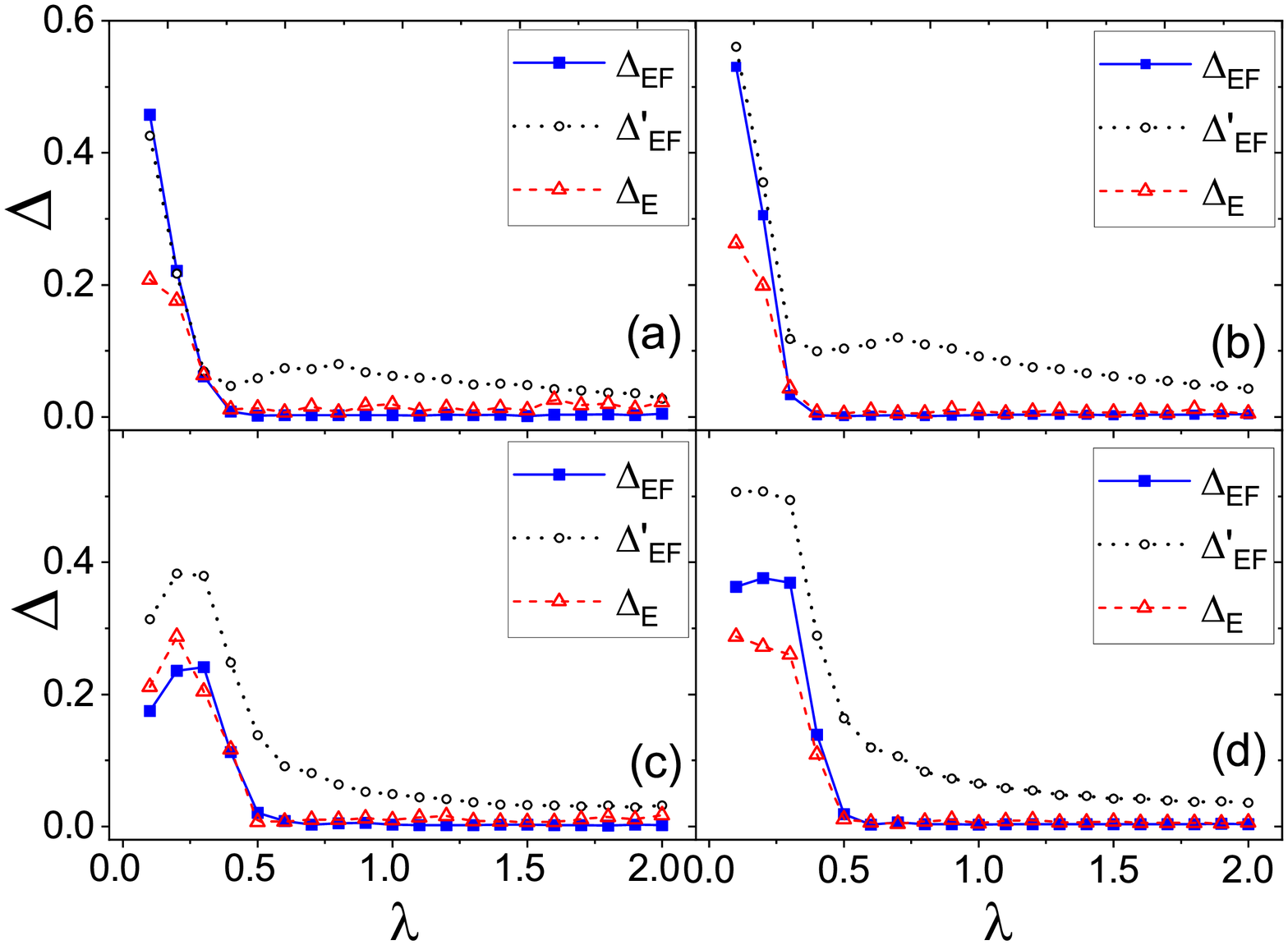}
\caption{``Distance'' to chaos in the LMG model and the Dicke model.
 The measures $\Delta_{EF}$ (solid squares) in Eq.(\ref{eq-DEF})
 and $\Delta' _{EF}$ (open circles) in Eq.(\ref{eq-doef}) are computed from statistical properties
 of EFs and the measure $\Delta_{E}$ (open triangles) in Eq.(\ref{eq-DE}) is computed from the statistics of spectra.
 (a) The LMG model with $\Omega=80$,
 (b) The LMG model with $\Omega=1000$,
 (c) the Dicke model with $N=80$,
 and  (d) the Dicke model with $N=1000$.
 The effective Planck constants are given by $1/\Omega$ and $1/N$, respectively, in the two modes.
  The two measures $\Delta_{EF}$ and $\Delta_{E}$ give almost the same results for the ``distance'' to chaos,
  when the systems are not far from chaos.
}\label{fig-psdisfinal-DickeLMG}
\end{figure}

 For symmetric states, the operators $K_{rs}$ can be written
 in terms of bosonic creation and annihilation operators $b^\dagger _r$ and $b_r$ \cite{chaos_Xu},
\be
K_{rs}=b_{r}^{\dagger}b_{s},\quad K_{r0}=K_{0r}^{\dagger}=b_{r}^{\dagger}
\sqrt{\Omega-b_{1}^{\dagger}b_{1}-b_{2}^{\dagger}b_{2}}
\ee
for $r,s=1,2$.
 Under the transformation,
\begin{equation}
b_{r}^{\dagger}=\sqrt{\frac{\Omega}{2}}(q_{r}-ip_{r}),\ \ \ b_{r}=\sqrt{\frac{\Omega}{2}}(q_{r}+ip_{r})
\end{equation}
for $r=1,2$, it is easy to verify that $q_{r}$ and $p_{s}$ obey the following commutation relation,
\begin{equation}
[q_{r},p_{s}]=\frac{i}{\Omega}\delta_{rs}.
\end{equation}
 Hence, $1/\Omega$ plays the role of an effective Planck constant,
\be
\hbar_{\rm eff}=\frac{1}{\Omega}.
\ee
 It is straightforward to find that the classical counterpart of model
 has the following Hamiltonian \cite{pre-98,Meredith98},
\be
H(\boldsymbol{p},\boldsymbol{q})=H_0(\boldsymbol{p},\boldsymbol{q})
 + \lambda V(\boldsymbol{p},\boldsymbol{q}),
\ee
where
\begin{gather}
H_{0}(\boldsymbol{p},\boldsymbol{q})=\frac{\epsilon'_{1}}{2}(p_{1}^{2}+q_{1}^{2})+\frac{\epsilon'_{2}}{2}(p_{2}^{2}+q_{2}^{2}), \nonumber \\
V(\boldsymbol{p},\boldsymbol{q})=\mu'_{1}(q_{1}^{2}-p_{1}^{2})(1-G/2)+\mu'_{2}(q_{2}^{2}-p_{2}^{2})(1-G/2) \nonumber \\
+\frac{\mu'_{3}}{\sqrt{2}}[(q_{2}^{2}-p_{2}^{2})q_{1}-2q_{2}p_{1}p_{2}]\sqrt{1-G/2} \nonumber \\
+\frac{\mu'_{4}}{\sqrt{2}}[(q_{1}^{2}-p_{1}^{2})q_{2}-2q_{1}p_{1}p_{2}]\sqrt{1-G/2},
\end{gather}
with $G=q_{1}^{2}+p_{1}^{2}+q_{2}^{2}+p_{2}^{2}\le2$.
 Here,  $\epsilon'_{1}=\epsilon_{1} \Omega, \epsilon'_{2}=\epsilon_{2} \Omega,
 \mu'_{1} = \mu_{1} \Omega^2, \mu'_{2} = \mu_{2} \Omega^2, \mu'_{3}= \mu_{3} \Omega^2$,
and $\mu_{4}'=\mu_{4} \Omega^2$.
In our numerical simulations, we set $\epsilon'_{1}=44.1,  \epsilon'_{2}= 64.5,
 \mu'_{1}= 62.1, \mu'_{2} = 70.2, \mu'_{3}= 76.5$, and $\mu'_{4} = 65.7$.
 Under this choice of the parameters,
 for a fixed value of $\lambda$, different values of $\Omega$ correspond to a same classical counterpart.

The second model is a single-mode Dicke model\cite{Dicke,Emary03},
which describes the interaction between a single bosonic mode and
a collection of $N$ two-level atoms. The system can be described
in terms of the collective operator ${\bf {J}}$ for the $N$
atoms, with
\begin{equation}
{J}_{z} = \sum_{i=1}^{N}{s}_{z}^{(i)},\ \ {J}_{\pm} = \sum_{i=1}^{N}{s}_{\pm}^{(i)},
\end{equation}
where ${s}_{\pm}^{(i)} = {s}_{x}^{(i)} \pm i {s}_{y}^{(i)}$ and
${s}_{x(y,z)}^{(i)}$ are Pauli matrices divided by $2$
for the $i$-th atom.
The Dicke Hamiltonian is written as~\cite{Emary03}
\begin{equation}
H=\omega_{0}J_{z}+\omega a^{\dagger}a+\frac{\lambda}{\sqrt{N}}\mu(a^{\dagger}+a)(J_{+}+J_{-}),
\end{equation}
 {which can also be written in the form of $H=H_0 + \lambda V$.}
The operator ${J}_{z}$ and ${J}_{\pm}$ obey the usual commutation
rules for the angular momentum,
\begin{equation}
[J_{z},J_{\pm}]=\pm J_{\pm},\ \ [J_{+},J_{-}]=2J_{z}.
\end{equation}

\begin{figure}
\includegraphics[width=1\linewidth]{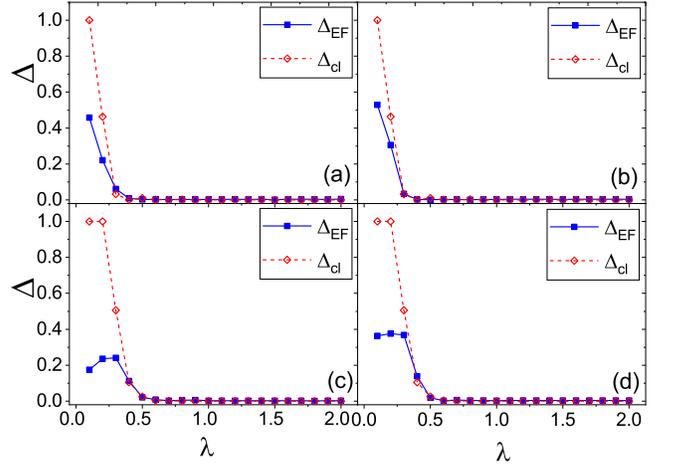}
\caption{``Distances'' to classical and quantum chaos in the LMG model and the Dicke model.
 The measure $\Delta_{EF}$ (solid squares) is the same as that in Fig.\ref{fig-psdisfinal-DickeLMG}.
 The measure $\Delta_{cl}$ (open diamonds)
 is defined in Eq.(\ref{CCL}) and was computed from properties of the corresponding classical phase spaces.
 (a) The LMG model with $\Omega=80$,
 (b) the LMG model with $\Omega=1000$,
 (c) the Dicke model with $N=80$,
 and  (d) the Dicke model with $N=1000$.
}\label{ChaosCL}
\end{figure}

The Hilbert space of this model is spanned by vectors $|j,m\rangle$ with $m=-j,-j+1,\cdots,j-1,j$,
 known as Dicke states.
 They are eigenstates of $\boldsymbol{J}^{2}$
and $J_{z}$, with $J_{z}|j,m\rangle=m|j,m\rangle$ and $\boldsymbol{J}^{2}|j,m\rangle=j(j+1)|j,m\rangle$.
 Below, we take $j$ as its maximal value, namely, $j=N/2$;
 it is a constant of motion, since $[\boldsymbol{J}^{2},H]=0$.
Another conserved observable in the Dicke model is the parity $\Pi$,
 given by $\Pi=\exp(i\pi\hat{N})$, where $\hat{N}=a^{\dagger}a+J_{z}+j$ is an operator for the ``excitation number'',
 counting the total number of excitation quanta in the system. In
our numerical study, we consider the subspace with $\Pi=+1$.

\begin{figure}
\includegraphics[width=1\linewidth]{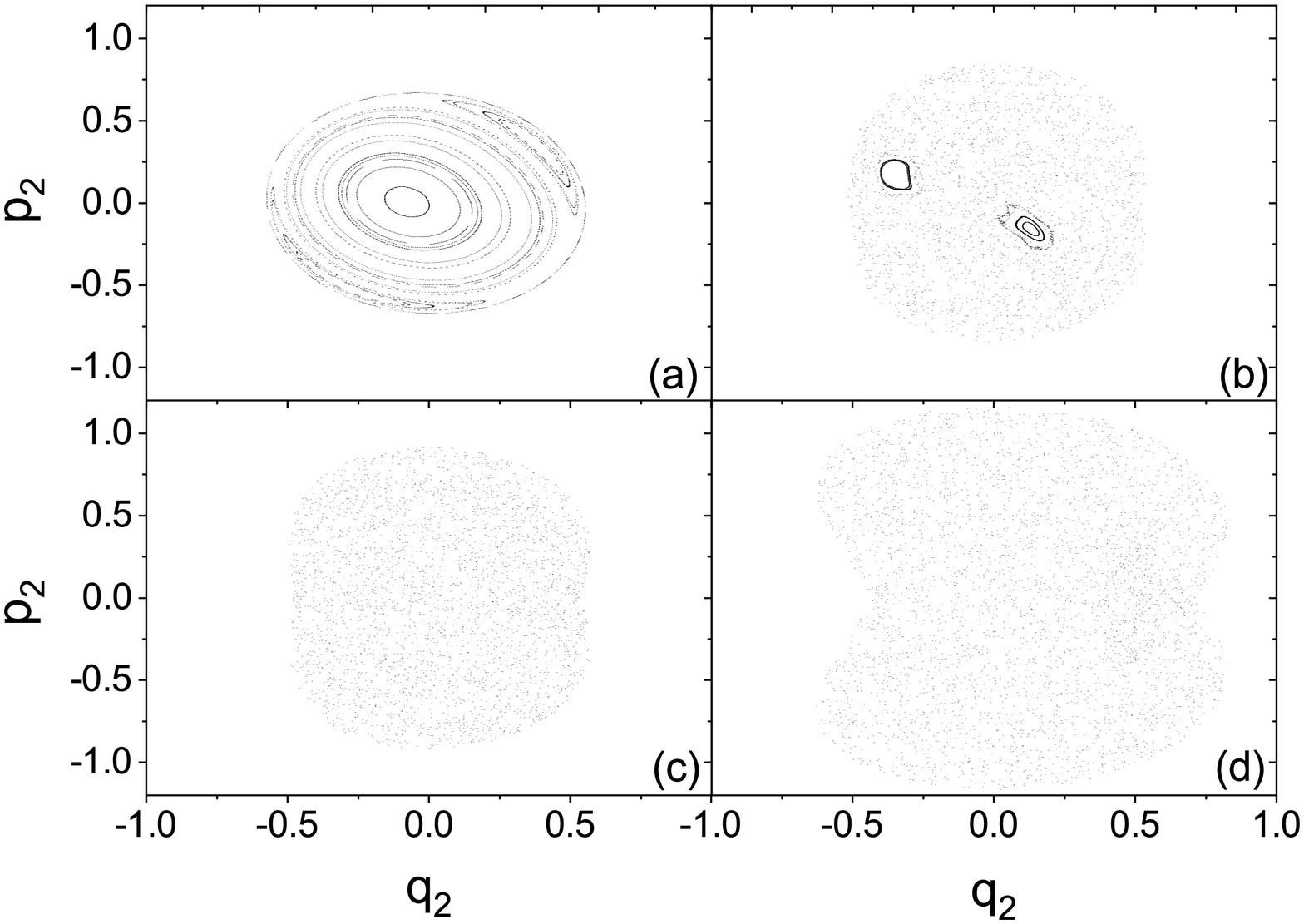}
\caption{Poincar\'{e} surfaces of section in the LMG model for $E=12+\sqrt{2}$ and $q_1=0$.
 (a) $\lambda=0.1$; (b) $\lambda=0.3$; (c) $\lambda=0.4$; (d) $\lambda=1.0$.
}\label{POSLMG}
\end{figure}

Making use of the Holstein-Primakoff representation of the angular
momentum operators,
\begin{gather}
J_{+}=b^{\dagger}\sqrt{2j-b^{\dagger}b},\ \ \ J_{-}=\sqrt{2j-b^{\dagger}b} \ b,\nonumber \\
J_{z}=(b^{\dagger}b-j),\label{eq-J}
\end{gather}
where $b$ and $b^\dag$ are bosonic operators satisfying $[b,b^{\dagger}]=1$,
the Hamiltonian can be further written in the following form,
\begin{gather}\label{}\notag
 H=\omega_{0}(b^{\dagger}b-j)+\omega a^{\dagger}a
 \\ +\lambda\mu(a^{\dagger}+a) \left(b^{\dagger}\sqrt{1-\frac{b^{\dagger}b}{2j}}+\sqrt{1-\frac{b^{\dagger}b}{2j}}b\right).
\end{gather}
We write Fock states related to $a^\dag$ and $b^\dag$ as $|n_{a}\rangle$
and $|n_{b}\rangle$, respectively,  for which
\begin{equation}
a^{\dagger}a|n_{a}\rangle=n_{a}|n_{a}\rangle,\ \ \ b^{\dagger}b|n_{b}\rangle=n_{b}|n_{b}\rangle.
\end{equation}
According to Eq.(\ref{eq-J}), $n_{b}$ should be truncated at $(n_{b})_{\rm max}=N$.
 In numerical simulations, we set $(n_{a})_{\rm max}=N$.
 Other parameters are $\omega_0=\omega=1/N$ and $\mu=1/N$.
 Under the transformation
\begin{gather}\begin{cases}
b^{\dagger}=\sqrt{\frac{N}{2}}(q_{1}-ip_{1}),\ \ \ b=\sqrt{\frac{N}{2}}(q_{1}+ip_{1}),\\
a^{\dagger}=\sqrt{\frac{N}{2}}(q_{2}-ip_{2}),\ \ \ a=\sqrt{\frac{N}{2}}(q_{2}+ip_{2}),
\end{cases}
\end{gather}
 one finds that
\begin{equation}
[q_{r},p_{s}]=\frac{i}{N}\delta_{rs}
\end{equation}
for $r=1,2$, and, hence, gets an effective Planck constant
\be
\hbar_{\rm eff}=\frac{1}{N}.
\ee
 The Hamiltonian of the classical counterpart of the model is written as
\be
H(\boldsymbol{p},\boldsymbol{q})=H_0(\boldsymbol{p},\boldsymbol{q})+\lambda V(\boldsymbol{p},\boldsymbol{q}),
\ee
where
\begin{gather}
H_0(\boldsymbol{p},\boldsymbol{q})=\frac{1}{2}(q_1^{2}+p_{1}^{2}+q_2^{2}+p_{2}^{2}-1), \nonumber \\
V(\boldsymbol{p},\boldsymbol{q})=2 q_1 q_2\sqrt{1-\frac{q_1^{2}+p_{1}^{2}}{2}}.
\end{gather}

\begin{figure}
\includegraphics[width=1\linewidth]{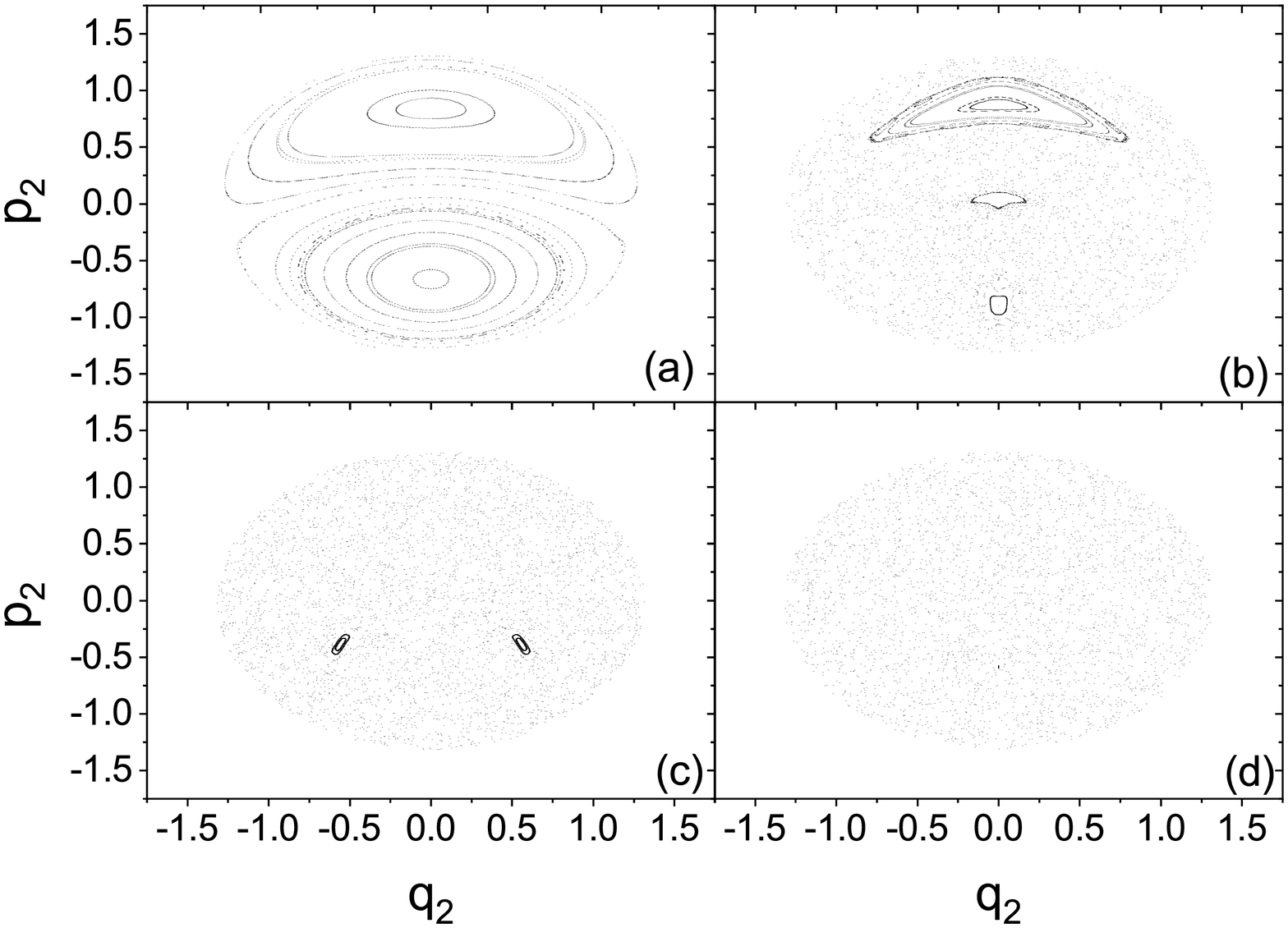}
\caption{Poincar\'{e} surfaces of section in the Dicke model for $E=\frac{\sqrt{3}-1}{2}$ and $q_1=0$.
 (a) $\lambda=0.2$; (b) $\lambda=0.4$; (c) $\lambda=0.6$; (d) $\lambda=1.0$.
}\label{POSDicke}
\end{figure}

\subsection{Numerical results}

In this section, we discuss results of numerical simulations performed in the LMG model and the Dicke model.
 We first test validity of the semiclassical result given in Eq.(\ref{eq-EFshapecl}).
 As seen in Fig.\ref{fig-EF2D-LMG} and Fig.\ref{fig-EF2D-Dicke},
 the average shape $\langle|C_{\alpha\boldsymbol{n}}|^{2}\rangle$ and
  its classical analog $\Pi(E_{\alpha},\boldsymbol{I_{n}})$ indeed show similar features in these two models.
 We have computed the difference between the two shapes, given by
\be
 d_c = \sum_{\boldsymbol n} |\langle |C_{\alpha \boldsymbol n}|^2\rangle - \Pi_N (E_\alpha, \boldsymbol{I_n})|,
\ee
 where $\Pi_N(E_\alpha, \boldsymbol{I_n})$ is the normalised $\Pi (E_\alpha, \boldsymbol{I_n})$.
 We found that $d_c=0.065$ in the LMG model and $d_c=0.08$ in the Dicke model.

 Then, we discuss properties of the distribution $f(R)$ for components $R_{\alpha \boldsymbol{n}}$
 in classically allowed regions with $\Pi(E_\alpha,\boldsymbol{I_n})\neq 0$.
 We found that this distribution is indeed quite close to the Gaussian form $f_G(R)$,
 when the underlying classical dynamics is chaotic, as illustrated in Fig.\ref{DisLMG} and Fig.\ref{DisDicke}
 with $\lambda=1$.
 In the computation of $f(R)$,  $100$ EFs in the middle energy region were used.
 The energy windows $\epsilon$ are as follows: In the LMG mode, $\epsilon\approx3$ for $\Omega=80$
 and $\epsilon\approx 0.02$ for $\Omega=1000$, in contrast to the total energy domain $\Delta E =64.5$
 in the unperturbed system;
 in the Dicke mode, $\epsilon\approx0.2$ for $N=80$ and $\epsilon\approx0.002$ for $N=1000$,
 in contrast to the total energy domain $\Delta E =2$.

\begin{figure}
\includegraphics[width=1\linewidth]{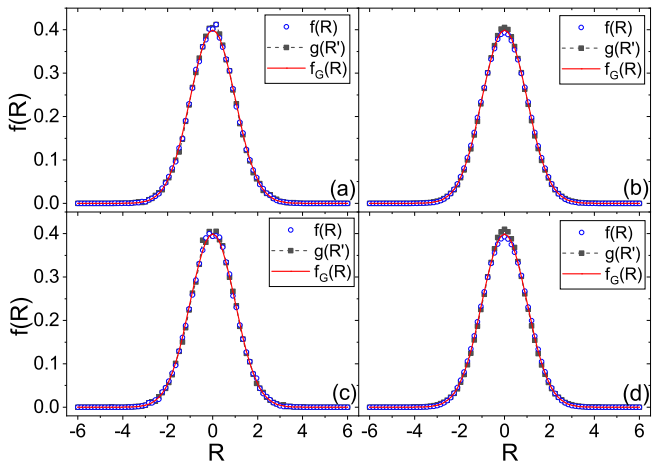}
\caption{The distribution $f(R)$ (open circles) and $g(R')$ (solid blocks with dashed lines) for $\lambda=0.5$
 in the defect Ising model and the defect XXZ model.
 The solid curve indicates the Gaussian distribution $f_{G}(R)$.
 (a) The defect Ising model with $N=10$; (b) the defect Ising model with $N=15$;
 (c) the defect XXZ model with $N=12$ and $S_z=-1$; (d) the defect XXZ model with $N=19$ and $S_z=-3.5$.
}\label{DischaosQ}
\end{figure}

 For comparison, we have also computed the distribution $g(R')$ given by another
 rescaling procedure, in which an average over unperturbed states is also performed
 (see discussions given in Sec.\ref{sect-measure}).
 In this rescaling procedure, as discussed in Refs.\cite{EFchaos-Benet03,EFchaos-Benet00},
 the average shape of EFs is expected to have the following semiclassical approximation,
\be\label{eq-efe}
 \langle|C_{\alpha \boldsymbol{n}}|^{2}\rangle' \simeq  \frac{S(E,E_{0})}{(2\pi\hbar)^{f}\rho(E)\rho_{0}(E_{0})},
\ee
 where $\rho_0(E_0)$ and $\rho(E)$ are the density of states of the two systems $H_0$
 and $H$, respectively,
 and $S(E,E_{0})$ indicates the overlap of the perturbed energy surface of $H=E$ and
 the unperturbed energy surface of $H_0=E_0$,
\be
S(E,E_{0})=\int d\boldsymbol{q}d\boldsymbol{p}\delta(E-H(\boldsymbol{p},\boldsymbol{q}))
\delta(E_{0}-H_{0}(\boldsymbol{p},\boldsymbol{q})).
\ee
The difference between $\langle |C_{\alpha \boldsymbol{n}}|^2\rangle'$ in Eq.(\ref{eq-efe}) and
$\langle |C_{\alpha \boldsymbol{n}}|^2\rangle$ in Eq.(\ref{eq-EFshapecl}) is quite clear.

In the computation of $g(R')$, only those rescaled components $R_{\alpha \boldsymbol{n}}'$
 in the region with nonzero $S(E_\alpha,E_{\boldsymbol{n}})$ were used.
 We found that, unlike the case of $f(R)$ discussed above,
 the distribution $g(R')$ usually shows obvious deviation from $f_G(R')$
 when the classical system is in the chaotic regime (Fig.\ref{DisLMG} and Fig.\ref{DisDicke}).
 Here, in the additional average for the unperturbed system, $100$ EFs in the middle energy region were used,
 with $\epsilon_0\approx 0.645$ in the LMG model and $\epsilon_0\approx0.02$ in the Dicke model.

 Variation of the measure $\Delta_{EF}$ in Eq.(\ref{eq-DEF})
 with the controlling parameter $\lambda$ is given in Fig.\ref{fig-psdisfinal-DickeLMG},
 together with the often-used
 measure given by $\Delta_E$ of the statistics of spectra.
 In order to improve the statistics, for each value of $\lambda$ , we used
 data obtained from several values of $\lambda'$ in a neighborhood of $\lambda$,
 $\lambda' \in [\lambda-0.05,\lambda+0.05]$.

 The agreement of the two measures $\Delta_{EF}$ and $\Delta_{E}$
 is already good in the case of not quite large $\Omega$ in the LMG model
 [Fig.\ref{fig-psdisfinal-DickeLMG}(a) with $\Omega=80$].
 The agreement becomes better, when the value of $\Omega$ is
 increased such that the system becomes closer to its classical limit [Fig.\ref{fig-psdisfinal-DickeLMG}(b)].
 Similar results were also found in the Dicke model [Fig.\ref{fig-psdisfinal-DickeLMG}(c) and (d)].
 Therefore, in these two models, the difference $\Delta_{EF}$ can be regarded
 as a good measure for the ``distance'' to chaos.

 For comparison, we have also computed the difference $\Delta' _{EF}$ given by
 the other rescaling procedure,
\be\label{eq-doef}
\Delta_{EF}'=\int|I_{g}(R')-I_{f_G}(R')|dR',
\ee
 where $I_{g}(R')$ denotes the cumulative distribution for $g(R')$.
 Due to the obvious difference between the distribution $g(R')$ and the Gaussian distribution
 shown in Fig.\ref{DisLMG} and Fig.\ref{DisDicke}, one expects a notable difference between
 $\Delta_{EF}'$  and $\Delta_E$.
 Indeed, as shown in Fig.\ref{fig-psdisfinal-DickeLMG}, unlike the case with $\Delta_{EF}$ discussed above,
 the agreement between $\Delta_{EF}'$  and $\Delta_E$ is not good.

 We have also computed a ``distance'' to chaos in the classical counterparts, denoted by $\Delta_{cl}$,
 which measures the proportion of regular region in energy surface.
 The measure is defined by
\be\label{CCL}
\Delta_{cl} = \lim_{N_T\to \infty} \frac{N_R}{N_T},
\ee
 where $N_T$ is a total number of points taken randomly in an energy surface of interest
 and $N_R$ is the number of the points for which $\lambda_L < \lambda_m$.
 Here, $\lambda_m$ is some small quantity and $\lambda_L$ is the Lyapunov exponent,
 defined as follows in the long time limit,
 \be
 \lambda_{L}=\lim_{t\rightarrow\infty}\lim_{d_{0}\rightarrow0}\frac{1}{t}\ln\frac{|d_{t}|}{|d_{0}|},
 \ee
 where $d_0$ denotes the initial phase-space distance and $d_t$ denotes the distance at a time $t$.
In our numerical simulation, we took  $t=1000$, $N_T=5000$, $\lambda_m=0.02$ in the LMG model
and $t=50000$, $N_T=5000$, $\lambda_m=0.001$ in the Dicke model.
 In Fig.\ref{ChaosCL}, it is seen that the agreement between the distances to quantum and classical
 chaos, characterized by $\Delta_{EF}$ and $\Delta_{cl}$, respectively, is quite good.
 Some examples of Poincar\'e surfaces of sections in the two models are shown in Fig.\ref{POSLMG} and Fig.\ref{POSDicke}.

\section{Numerical simulations in models without classical counterparts}\label{sect-num-Ising}

 In this section, we study the distribution of rescaled components of EFs in
 models without any classical counterpart.
 It seems reasonable to expect that the final result of Sec.\ref{sect-chaosEF-sa}, that is, that the distribution of
 appropriately rescaled
 components should have a Gaussian form, may be valid to some extent in this type of models as well.

\begin{figure}
\includegraphics[width=0.96\linewidth]{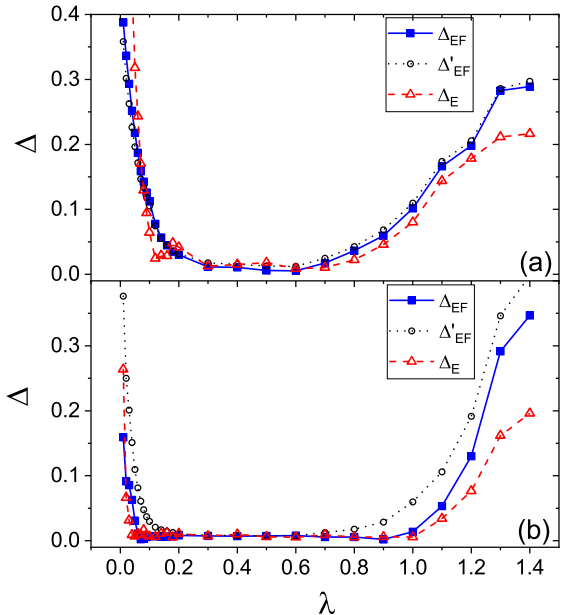}
 \caption{Similar to Fig.\ref{fig-psdisfinal-DickeLMG}, but for the defect Ising model with (a) $N=10$ and (b)$N=15$.
}\label{DEF-DI}
\end{figure}

 Here, a major problem met is the determination of the region of components that should be taken into account.
 As discussed previously, in a system with a classical counterpart,
 this region corresponds to the classically energetically allowed region.
 For a system without any classical counterpart,  this is a highly nontrivial problem.
 In this paper, we do not intend to solve this problem, but, to
 circumvent it by restricting ourselves to models, whose EFs occupy almost the whole unperturbed
 energy region.
 In this type of models, one can simply use all components of the EFs when computing $f(R)$.
 Specifically, we study a defect XXZ model and a defect Ising model,
 adopting a periodic boundary condition in numerical simulations.

 The defect XXZ model \cite{d-XXZ} is a modified XXZ model, in which an external magnetic field is
 applied on two sites of the $N$ spins.
The unperturbed Hamiltonian and the perturbation have the following forms,
\begin{gather}
H_{0}=\sum_{i=1}^{N}s_{x}^{i}s_{x}^{i+1}+s_{y}^{i}s_{y}^{i+1}
+ \mu_{z} \sum_{i=1}^{N}s_{z}^{i}s_{z}^{i+1}
 \\  V=\mu_{1}s_{z}^{1}+\mu_{4}s_{z}^{4},
\end{gather}
 where the periodic boundary condition implies that
 $s^{N+1}_{a}=s^1_{a}$ for $a=x,y,z$.
 The system is a quantum chaotic system for $\lambda$ within an appropriate regime, while,
 it exhibits the so-called many-body localisation for $\lambda$ sufficiently large.
 The Hamiltonian $H$ is commutable with $S_{z}$, the $z$-component of the total
spin, and we consider a subspace with a definite value of $S_{z}$ in our numerical
study. Other parameters used in this model are $\mu_{1}=1.11$, $\mu_{4}=1.61$,
and $\mu_{z}=1$.

 The defect Ising model is a transverse Ising model, in which an additional magnetic field is
 applied on two sites of the $N$ spins, with
\begin{gather}
H_{0}=\sum_{i}^{N}s_{z}^{i}s_{z}^{i+1}+\mu_{x}\sum_{i=1}^{N}s_{x}^{i}, \\
V=\mu_{1}s_{z}^{1}+\mu_{4}s_{z}^{4}.
\end{gather}
Similarly, it is a quantum chaotic system for $\lambda$ in an appropriate regime and exhibits
 many-body localisation for $\lambda$ sufficiently large. The parameters used
are $\mu_{1}=1.11$, $\mu_{4}=1.61$, and $\mu_{x}=0.6$.

\begin{figure}
\includegraphics[width=0.96\linewidth]{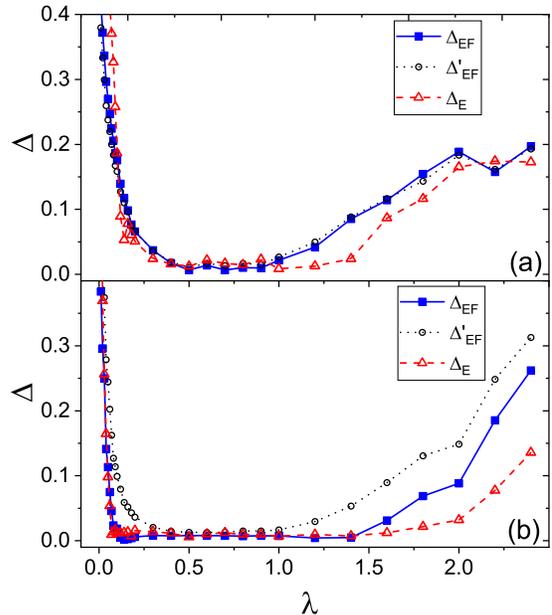}
 \caption{Similar to Fig.\ref{fig-psdisfinal-DickeLMG}, but for the defect XXZ model with (a) $N=12$, $S_z=-1$,
  and (b)$N=19$, $S_z=-3.5$.
}\label{DEF-DX}
\end{figure}

 Our numerical simulations reveal that, in these two models,
 the distributions $f(R)$ are also quite close to the Gaussian form $f_G(R)$,
 when the statistics of the spectra is close to the prediction of RMT
 as illustrated in Fig.\ref{DischaosQ} with $\lambda=0.5$.
 Unlike the two models discussed in the previous section, the distributions
 $g(R')$ are also close to the Gaussian form at $\lambda =0.5$.

 In the computation of $f(R)$,  $50$ EFs in the middle energy region were used.
 The energy windows $\epsilon$ are as follows: In the defect Ising model,
 $\epsilon\approx 0.2$ and $\epsilon_0=0.02$ for $N=10$
 in contrast to the total energy domain $\Delta E =7.11$,
 and $\epsilon\approx 0.07$ for $N=15$ in contrast to $\Delta E = 10.64$;
 in the defect XXZ model, $\epsilon\approx0.3$ and $\epsilon_0=0.02$ for $N=12$, $S_z=-1$ in contrast to $\Delta E = 8.03$,
 and $\epsilon\approx0.01$ for $N=19$, $S_z=-3.5$
 in contrast to the total energy domain $\Delta E =10.66$.

\begin{figure}
\includegraphics[width=0.98\linewidth]{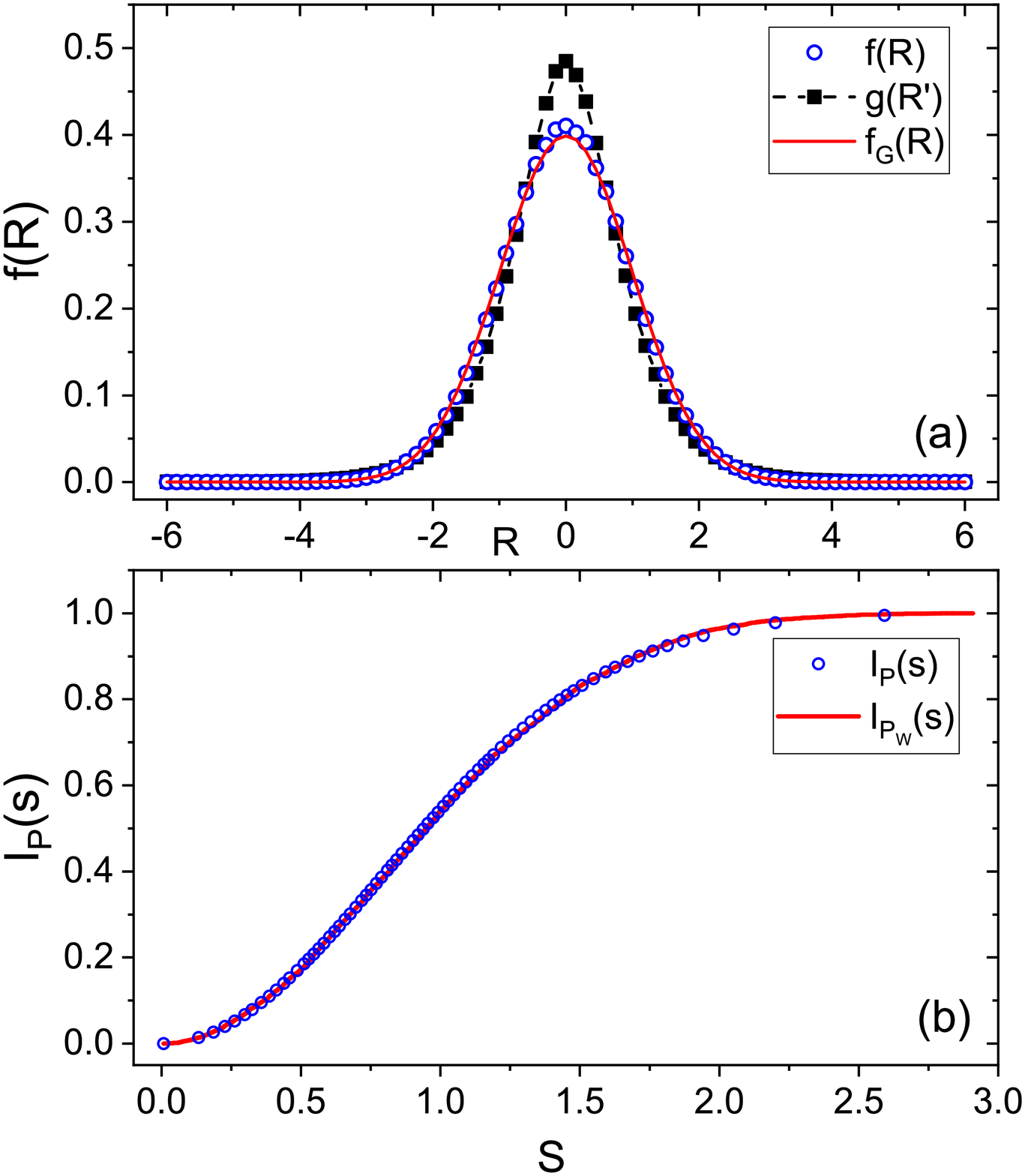}
\caption{(a) The distributions of $f(R)$ (open circles) and $g(R')$
(solid blocks with dashed lines) in the defect Ising model with $N=15$ and $\lambda=0.06$.
(b) The corresponding cumulative distribution of the nearest-level-spacing
 distribution (open circles).
 The solid curve indicates the cumulative distribution given by the Wigner surmise.
}\label{DisW-DI}
\end{figure}

 The two measures $\Delta_{EF}$ and $\Delta_E$ exhibit similar behaviors,
 like the cases discussed in the previous section for the LMG and Dicke models
 (Fig.\ref{DEF-DI} and Fig.\ref{DEF-DX}).
 Thus, at least in these two models, the difference $\Delta_{EF}$ can be
 regarded as a good measure for the ``distance'' to chaos.

 In consistence with the behaviors of the distribution $g(R')$ illustrated in Fig.\ref{DischaosQ},
 the two quantities $\Delta'_{EF}$ and $\Delta_{EF}$ are close in most regions
 where the systems are chaotic systems according to their spectra statistics.
 That is, in most cases, an average over the unperturbed energy does not bring much difference
 in the defect Ising and defect XXZ models.
 This may be partially related to the fact that EFs in these two models occupy almost the whole energy region
 for $\lambda $ not small.

\begin{figure}
\includegraphics[width=0.98\linewidth]{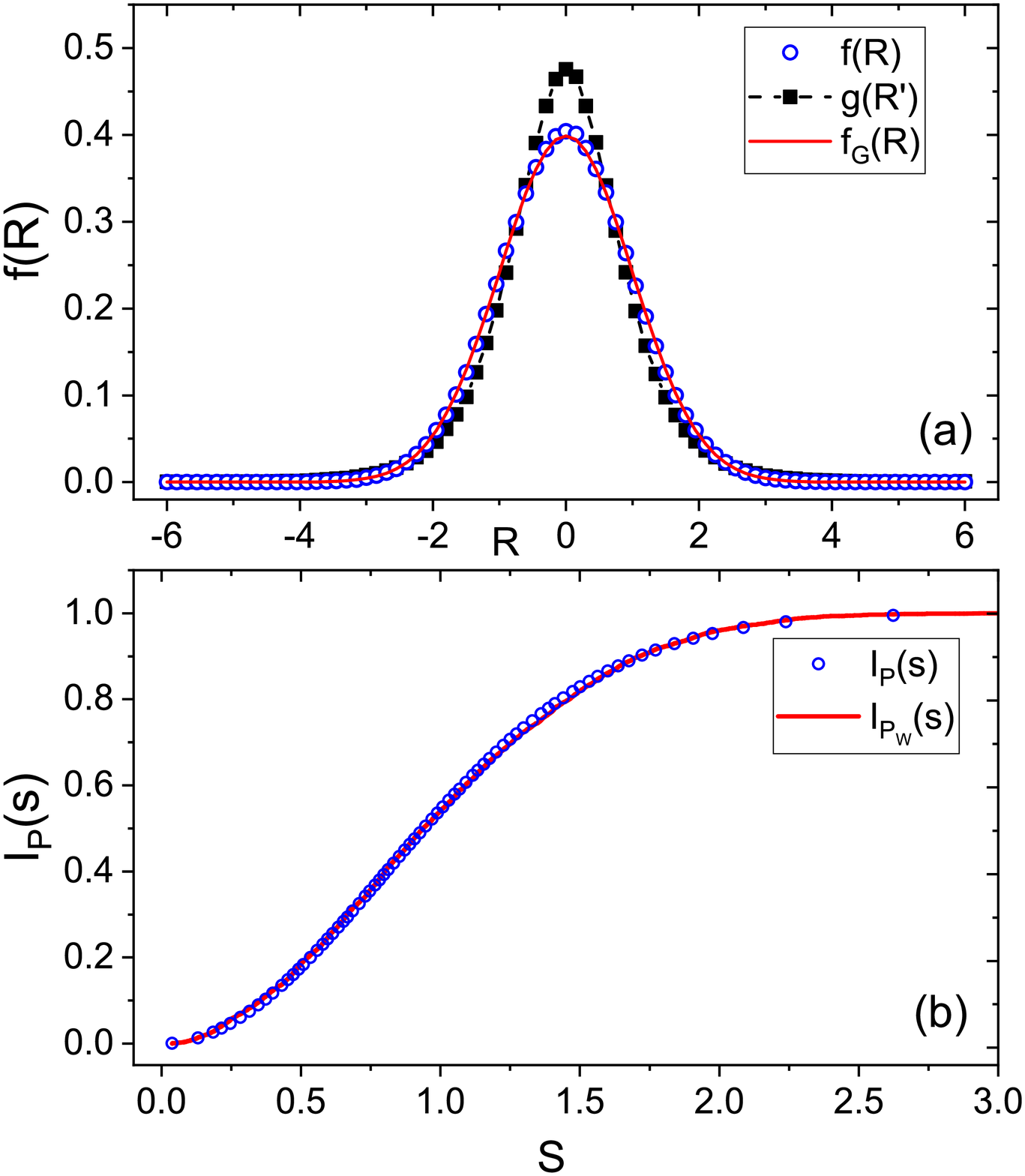}
\caption{Similar to Fig.\ref{DisW-DI}, but for the defect XXZ model with $N=19$, $S_z=-7$ and $\lambda=0.12$.
}\label{DisW-DX}
\end{figure}

 There are still some regions of $\lambda$ in Fig.\ref{DEF-DI}(b) and Fig.\ref{DEF-DX}(b)
 with relatively large Hilbert spaces,
 in which $\Delta'_{EF}$ shows some notable deviation from $\Delta_{EF}$ and $\Delta_{E}$.
 Some examples of the distributions $f(R)$ and $g(R')$ in this case
 are shown in Fig.\ref{DisW-DI} and Fig.\ref{DisW-DX}, together with the corresponding
 distributions of $I_P(s)$ and $I_{P_W}(s)$.

\section{Conclusions}\label{sect-Conclusion}

 In this paper, based on semiclassical analysis,
 it has been shown that those components of EFs of quantum chaotic systems, 
 which lie in classically-allowed regions
 of integrable bases, can be regarded as random numbers in a sense
 similar to that stated in the Berry's conjecture.
 For the distribution $f(R)$ of these components to have a Gaussian form,
 which is predicted by the RMT,  an appropriated rescaling procedure
 with respect to the average shape of EFs is needed,
 where the average should be taken over perturbed states with neighbouring energies.
 It is found that an additional average over unperturbed basis states with neighbouring unperturbed energies
 may cause deviation of the distribution of rescaled components of EFs from the Gaussian form.

 The above results suggest that deviation of the distribution $f(R)$ from the Gaussian
 distribution may be used as a measure for the ``distance'' to quantum chaos.
 In two models possessing classical counterparts, when the perturbed system goes from
 integrable to chaotic with the increase of perturbation strength,
 our numerical simulations show that this deviation
 coincides with the deviation of the nearest-level-spacing
 distribution from the prediction of RMT.

 It is known that specific dynamics of the underlying classical systems
 may induce certain modifications to the Berry's conjecture
 \cite{KpHl,Heller87,Bies01,Sr96,Sr98,Backer02,Urb03,Kp05}.
 Since the main result of this paper is based on this conjecture,
 specific underlying classical dynamics may have some
 influence in results of this paper as well.
 In particular, it may induce some deviation of the distribution $f(R)$ for some EFs
 from the Gaussian distribution.
 However, if sufficiently many EFs are used in the computation of
 $f(R)$, it is reasonable to expect that the induced deviation should be small.

 In two models without simple classical counterpart,
 we have found similar numerical results about the distribution of $f(R)$.
 Analytical explanation of this point is still lacking.
 It seems that the following feature of these two models may be of relevance.
 That is, in both models the matrices of the perturbations $V$ in the unperturbed bases do not have
 a clear band structure; in other words, the perturbation couples basis vectors
 far separated in energy.
 We hope that these numerical results may stimulate more investigations.

\acknowledgements
 This work was partially supported by the Natural Science Foundation of China under Grant
 Nos.~11275179, 11535011, and 11775210.


\begin{thebibliography}{99}
\bibitem{foot-region} The distribution of rescaled components computed in Ref.\cite{EFchaos-WW}
 is for the so-called nonperturbative
 regions of EFs. The relation between these regions and classically-allowed regions
 will be discussed in a different paper.
\end{thebibliography}
\end{document}